\documentclass{WileyMSP-template}
\usepackage{xcolor}
\usepackage{siunitx}
\usepackage{amsmath}

\begin{document}

\pagestyle{fancy}


\title{Curvature-Controlled Polarization in Adaptive Ferroelectric Membranes}

\maketitle


\author{Greta Segantini*}$^{1}$
\author{Ludovica Tovaglieri*}$^{1}$
\author{Chang Jae Roh*}$^{1}$
\author{Chih-Ying Hsu}$^{1,2}$
\author{Seongwoo Cho}$^{1}$
\author{Ralph Bulanadi}$^{1}$
\author{Petr Ondrejkovic}$^{3}$
\author{Pavel Marton}$^{4,3}$
\author{Jirka Hlinka}$^{3}$
\author{Stefano Gariglio}$^{1}$
\author{Duncan T.L. Alexander}$^{2}$
\author{Patrycja Paruch}$^{1}$
\author{Jean-Marc Triscone}$^{1}$
\author{Céline Lichtensteiger}$^{1}$
\author{Andrea D. Caviglia}$^{1}$

\begin{affiliations}
(1)Department of Quantum Matter Physics, University of Geneva, 24 Quai Ernest-Ansermet, CH-1211 Geneva 4, Switzerland.\\

(2)Electron Spectrometry and Microscopy Laboratory (LSME), Institute of Physics (IPHYS), Ecole Polytechnique Fédérale de Lausanne (EPFL), CH-1015 Lausanne, Switzerland.\\

(3)Institute of Physics of the Czech Academy of Sciences, Na Slovance 2, Praha 8, 182 00, Czech Republic.\\

(4)Institute of Mechatronics and Computer Engineering, Technical University of Liberec, Studentska 2, Liberec, 46117, Czech Republic.\\

\end{affiliations}


\keywords{Oxide Membranes, Ferroelectrics, Strain, Ferroelectric Domains, Flexible Electronics}\\
\bigskip
*These authors have equally contributed to the manuscript.

\justify
\begin{abstract}

In this study, we explore the ferroelectric domain structure and mechanical properties of PbTiO$_3$-based membranes, which develops a well-ordered and crystallographic-oriented ripple pattern upon release from their growth substrate. The ferrolectric domain structure of the PbTiO$_3$ layer was examined at various length scales using optical second harmonic generation, piezoresponse force microscopy, and scanning transmission electron microscopy. These methods reveal the presence of purely in-plane domains organized into superdomains at the crest of the ripples, while an in-plane/out-of-plane domain structure was observed in the flat regions separating the ripples, in agreement with phase-field simulations. The mechanical properties of the membrane were assessed using contact resonance force microscopy, which identified distinct mechanical behaviors at the ripples compared to the flat regions. This study shows that the physical properties of the ferroelectric layer in membranes can be locally controlled within an ordered array of ripples, with well-defined geometric characteristics.

\end{abstract}

\section{Introduction}

Epitaxial growth in ferroelectric thin films allows precise two-fold control over domain configurations.
Firstly, by carefully selecting substrates with specific symmetry and lattice parameters, the strain applied to the ferroelectric film can be tuned to induce complex, well-defined domain patterns \cite{Damodaran2016,schlom2007strain}. Secondly, electrostatic boundary conditions, established via interface and surface electrodes or dielectric spacer layers, are equally crucial in shaping and stabilizing domain structures \cite{Lichtensteiger-NanoLett-2014,Cazorla2015,Lichtensteiger-NJP-2016}.
Such control over domains is a powerful tool for tailoring the physical and functional properties of ferroelectric materials \cite{Jang2009,Zubko-Ferroelectrics-2012,Zubko-Nature-2016,Strkalj2019,nataf2020domain,Gradauskaite2022}. For example, their piezoelectric response can be significantly enhanced by modulating the size, orientation, and arrangement of domains \cite{Eng1999,wada2004enhanced,hong2014piezoelectric,Shi2025}. The resulting domain walls also play a critical role, not only influencing domain stability and switching dynamics \cite{PARUCH2013667,Meier2021}, but also conferring their own emergent properties, sometimes absent in the parent material \cite{catalan2012domain}.
Additionally, ferroelectric domain configurations can impact the structural properties of adjacent epitaxial thin films through mechanical coupling \cite{Lichtensteiger2023}. 
In some cases, ferroelectric domains can self-organize into superdomains, which are highly ordered structures that introduce periodicity and hierarchical order \cite{Matzen2014,Braun-Nanotechnology-2018}. This self-organization enhances ferroelectric functionality by influencing macroscopic properties such as polarization switching, dielectric response, and electromechanical coupling \cite{Scott2017}.  
Domain, superdomain, and domain wall engineering are essential for controlling the functional properties of ferroelectrics for device applications across various technological areas \cite{Li2024, Lipatov2019}.

While epitaxial growth enables the formation of complex domain structures, it often limits the achievable configurations to those compatible with the strain imposed by the growth substrate. 
The development of freestanding ferroelectric oxide membranes using sacrificial layers \cite{Lu2016} offers a unique opportunity to achieve domain configurations going beyond the constraints of epitaxy \cite{Dong2019,Han2020,Guo2020,Cai2022}.
Upon detaching from the growth substrate, the ferroelectric material experiences changes in both electrostatic and mechanical conditions, resulting in unique deformations such as ripples, wrinkles, and buckles. These deformations can be characterized by complex domain structures and exhibit enhanced flexibility. Ferroelectric oxide membranes present unique states, which could unlock novel functional properties with significant potential for next-generation electronic devices, including flexible electronics \cite{Dong2020,Sun2022,Han2022,Yu2023,pesquera2024hierarchicaldomainstructuresbuckled}.

In this work, we investigate the domain structure and mechanical properties of a PbTiO$_3$ thin film deposited on a SrRuO$_3$ bottom electrode, following the release of the heterostructure from its growth substrate. 
Upon release, the membrane exhibits a well-ordered, adaptive ripple pattern, marking, to the best of our knowledge, the first observation of such a long-range periodic arrangement in a freestanding ferroelectric system.
We study the domain structure of the membrane at different length scales using optical second harmonic generation (SHG), piezoresponse force microscopy (PFM), and scanning transmission electron microscopy (STEM). These techniques reveal the presence of purely in-plane domains ($a$ domains), organized into superdomains at the top of the ripples, with an in-plane/out-of-plane ($a/c$) domain structure observed elsewhere. The mechanical properties of the membrane were investigated using contact-resonance force microscopy (CRFM), which showed distinct mechanical properties at the ripples compared to the flat regions.

\section{Results and Discussion}

\begin{figure}[ht!]
\centering
  \includegraphics[width=\textwidth]{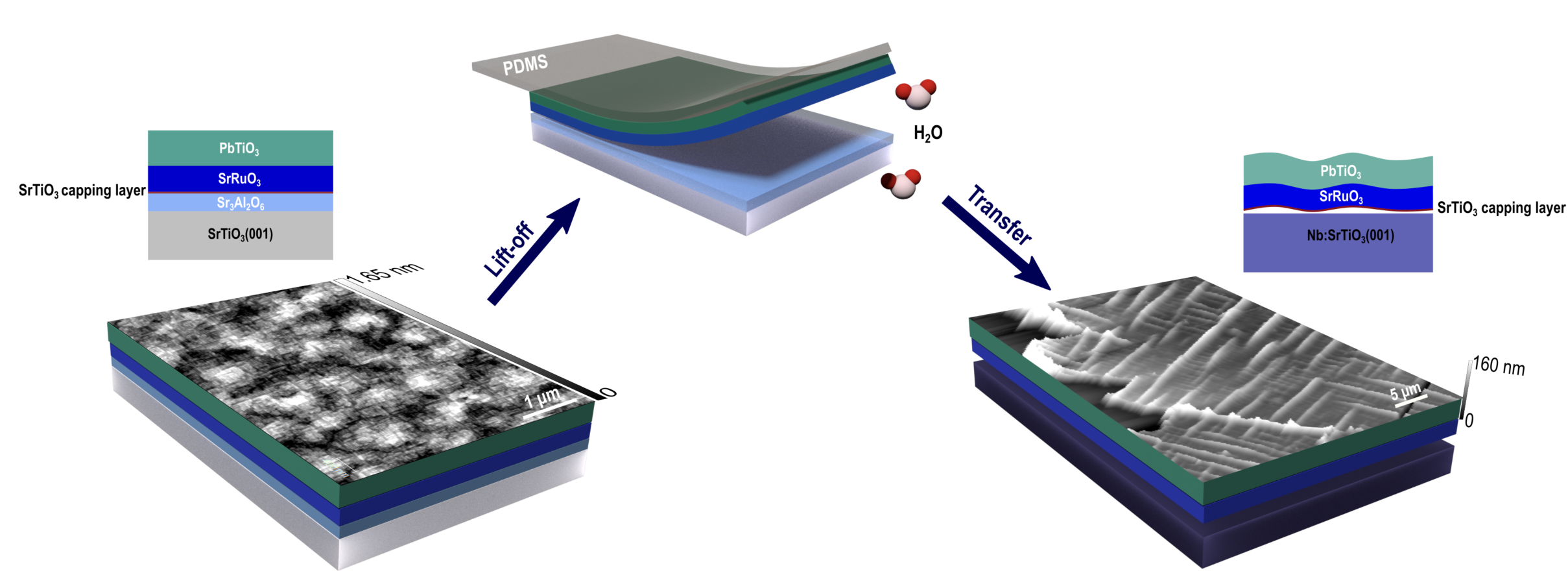}
  \caption{Schematic of the fabrication process for the PbTiO$_3$/SrRuO$_3$ membrane. From left to right: the as-grown PbTiO$_3$/SrRuO$_3$/SrTiO$_3$/Sr$_3$Al$_2$O$_6$/SrTiO$_3$(001) heterostructure displayed with the AFM topography image showing a flat surface with an RMS roughness of approximately 0.4 nm. The lift-off procedure, where the Sr$_3$Al$_2$O$_6$ sacrificial layer dissolves in deionized water, leaving the membrane attached to a PDMS support. Finally, the membrane is transferred onto a Nb:SrTiO$_3$(001)-oriented target substrate. The overlaid three-dimensional (3D) AFM topography image clearly reveal the formed ripple pattern. (The AFM topography images are also provided in SI Figures S1a and S1b.)}
\label{fig:fab_process}
\end{figure}

The fabrication of the PbTiO$_3$-based membrane involves several steps, including high-quality controlled epitaxial growth of an oxide hetrostructure, as well as the precise separation and transfer of the membrane. A 15 nm-thick sacrificial Sr$_3$Al$_2$O$_6$ layer was grown on a SrTiO$_3$(001)-oriented substrate using pulsed laser deposition (PLD). A 4-unit-cell-thick SrTiO$_3$ capping layer was deposited by PLD on the Sr$_3$Al$_2$O$_6$ layer to prevent degradation upon exposure to air. 
The SrRuO$_3$ and the PbTiO$_3$ layers were subsequently deposited via off-axis radio frequency (RF) magnetron sputtering. 
After the lift-off of the sacrificial layer, the released membrane consisted of PbTiO$_3$(30 nm)/SrRuO$_3$(22nm)/SrTiO$_3$(1.6 nm). In the remainder of the manuscript, we will refer to this membrane simply as the PbTiO$_3$/SrRuO$_3$ membrane.
\textbf{Figure \ref{fig:fab_process}} schematically illustrates the lift-off process and dissolution of the sacrificial layer, and the subsequent transfer of the membrane onto the niobium-doped SrTiO$_3$(001) (Nb:SrTiO$_3$(001)) target substrate. Detailed descriptions of the lift-off, transfer procedure, and layer growth conditions are provided in the Experimental section.
Atomic force microscopy (AFM) topography images overlaid in Figure \ref{fig:fab_process} highlight a flat surface, with a root mean square (RMS) roughness of approximately 0.4 nm, before the transfer (left), and the formation of organized ripples once the membrane is created (right).
The full topography images corresponding to Figure \ref{fig:fab_process} are provided in Supporting Information (SI) Figure S1, while SI Figure S2 presents optical images of the membrane, where the ordered ripples are clearly visible. 

Rippling phenomena have been previously observed in freestanding ferroelectric systems and are commonly linked to the redistribution of electrostatic forces and mechanical stresses following the release of the heterostructure from the growth substrates \cite{Li2022,Huang2024,pesquera2024hierarchicaldomainstructuresbuckled}. 
In this study, the release of the membrane leads to the formation of ripples that preferentially develop along the [100]$_{pc}$ and [010]$_{pc}$ axes of the PbTiO$_3$ film, representing the first observation of such an arrangement.
Throughout the manuscript, the directions of the crystallographic axes [100]$_{pc}$ and [010]$_{pc}$ are indicated for clarity, but note that they can be permuted due to the cubic symmetry of the SrTiO$_3$ substrate used for the growth.

\begin{figure}[ht!]
\centering
  \includegraphics[width=0.9\textwidth]{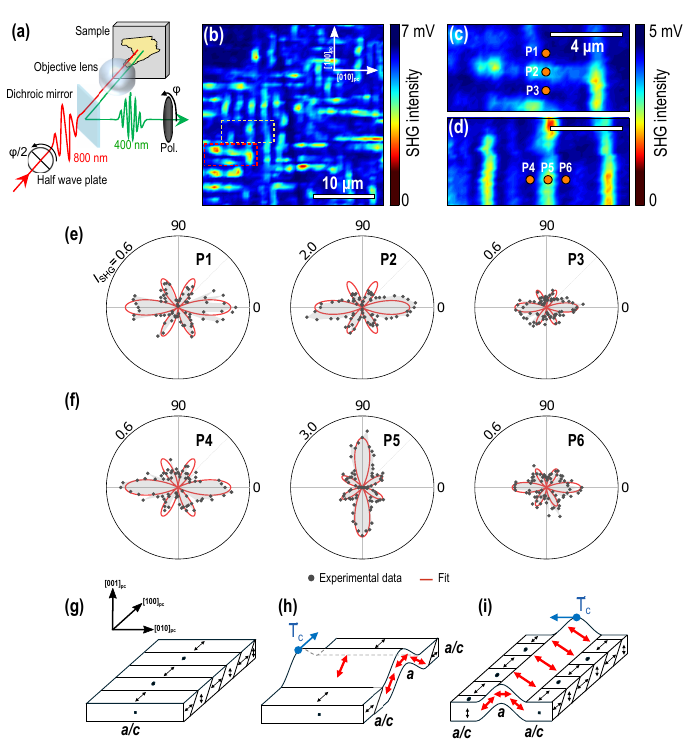}
  \caption{Ferroelectric symmetry evolution on the PbTiO$_3$/SrRuO$_3$ ripples. (a) Optical second harmonic generation (SHG) microscopy experiment. (b-d) 2D SHG intensity distributions on the PbTiO$_3$/SrRuO$_3$ membrane. (e-f) Polarization-angle dependence of SHG intensity as a function of position, crossing a ripple oriented along the [010]$_{pc}$ axis (e) and a ripple oriented along the [100]$_{pc}$ axis (f). (g-i) Schematic description of the ferroelectric polarization states on (g) flat region, (h) [010]$_{pc}$-oriented ripple, and (i) [100]$_{pc}$-oriented ripple. The red arrows and dots represent the polarization axis of each single domain. The blue arrows in (h) and (i) indicate the tangential direction to the ripple curvature.}
\label{fig:SHG}
\end{figure}

To characterize the domain structure of the membrane, a multi-scale approach was adopted, spanning from the micrometer down to the nanometer scale.
First, SHG microscopy, displayed in \textbf{Figure \ref{fig:SHG}a}, was employed. This is a sensitive technique for investigating inversion symmetry breaking and the associated ferroelectric properties in ferroelectric materials \cite{Cherifi-Hertel2017,Cherifi-Hertel-JAP-2021,roh2018deterministic,ju2021possible,jeong2024individual}. The SHG experiment was performed in normal incidence geometry to monitor the in-plane symmetry component of the polarization (see Experimental section). Before conducting a microscopic investigation of the ripples, the presence of a finite SHG response was confirmed across the entire PbTiO$_3$/SrRuO$_3$ membrane flake, while no SHG signal was detected outside the flake region (see SI Figure S4). Since the contribution of the SrRuO$_3$ layer to the SHG response is negligible \cite{chauleau2017multi,Roh2021}, the observed SHG signal is mainly attributed to the ferroelectric properties of the PbTiO$_3$ layer within the PbTiO$_3$/SrRuO$_3$ flake.

Figure \ref{fig:SHG}b presents a two-dimensional (2D) SHG intensity map over a 30$\times$30 $\mu$m$^2$ area, monitored with diffraction-limited resolution. 
Notably, the SHG image reveals an inhomogeneous intensity distribution with locally enhanced, periodic structures aligned along the [010]$_{pc}$ and [100]$_{pc}$ directions. A comparison with optical microscopy and surface topography images (see SI Figures S1b and S2) shows that the spatial distribution of SHG intensity enhancement closely matches the ripple arrangement in the PbTiO$_3$/SrRuO$_3$ membrane. This indicates that the SHG response effectively distinguishes the flat regions from the ripples within the membrane. In particular, the SHG intensity is enhanced by a factor of three to five on the ripples, suggesting that the structural symmetry is modified in these regions.

To gain deeper insight into the local symmetry, we characterize the SHG signal at specific positions and, in particular, trace its evolution when crossing the ripples. Figures \ref{fig:SHG}c and \ref{fig:SHG}d present the SHG intensity maps obtained with higher spatial resolution, which correspond to the areas highlighted by the yellow and red boxes in Figure \ref{fig:SHG}b. 
Focusing on two ripples with different orientations, six positions were selected: P1, P2, and P3 for the horizontal ripple, and P4, P5, and P6 for the vertical ripple, respectively. Figures \ref{fig:SHG}e and \ref{fig:SHG}f display the polarization-dependent SHG responses at these positions. When crossing the horizontal ripple from P1 to P3 (Figure \ref{fig:SHG}e), the SHG intensity increases approximately by three times at the ripple (P2) and returns to its original level on the opposite side (P3), while the SHG pattern remains unchanged. In contrast, when crossing the vertical ripple from P4 to P6 (Figure \ref{fig:SHG}f), both the SHG intensity and pattern undergo significant changes at the ripple (P5). The SHG pattern exhibits a precise 90$^\circ$ rotation, and the intensity increases by five times compared to the flat regions (P4 and P6).

To determine the crystal symmetry, the obtained SHG patterns were analyzed by considering the electric dipole contribution of the polar tetragonal structure with the point group $m4m$ (see SI section 1 and SI Figure S5a and Figure S5b). 
Additionally, SHG simulations reflecting ferroelectric domain states revealed that the two-fold SHG patterns observed in the flat regions (P1, P3, P4, and P6) correspond to the $a$ domain state component in the $a/c$ structure (see SI section  1 and SI Figure S5c and Figure S5d). 
Based on this result, the SHG patterns observed at the horizontal (P2) and vertical (P5) ripples were successfully fitted by incorporating larger nonlinear susceptibility tensors and a 90$^\circ$-rotated broken symmetry state. 

Considering that the $a/c$ domain structure stabilizes in the flat regions, as it typically occurs in tetragonal PbTiO$_3$ (Figure \ref{fig:SHG}g)\cite{Han2020,catalan2011flexoelectric}, the SHG results at the ripples suggest two key effects of the ripple orientation on the polarization states: (1) when the tangential direction ($T_c$) of a ripple curvature is parallel to the in-plane polarization state in the flat region (Figure \ref{fig:SHG}h), an enhancement of the polarization state is induced. (2) When the tangential direction is perpendicular to the polarization (Figure \ref{fig:SHG}i), both an enhancement and a rotation of the polarization state occur.
These findings indicate that the direction of a curved structure plays a crucial role in manipulating structural symmetry and its associated physical properties, such as ferroelectric behavior.

\begin{figure}[ht!]
\centering
  \includegraphics[width=\textwidth]{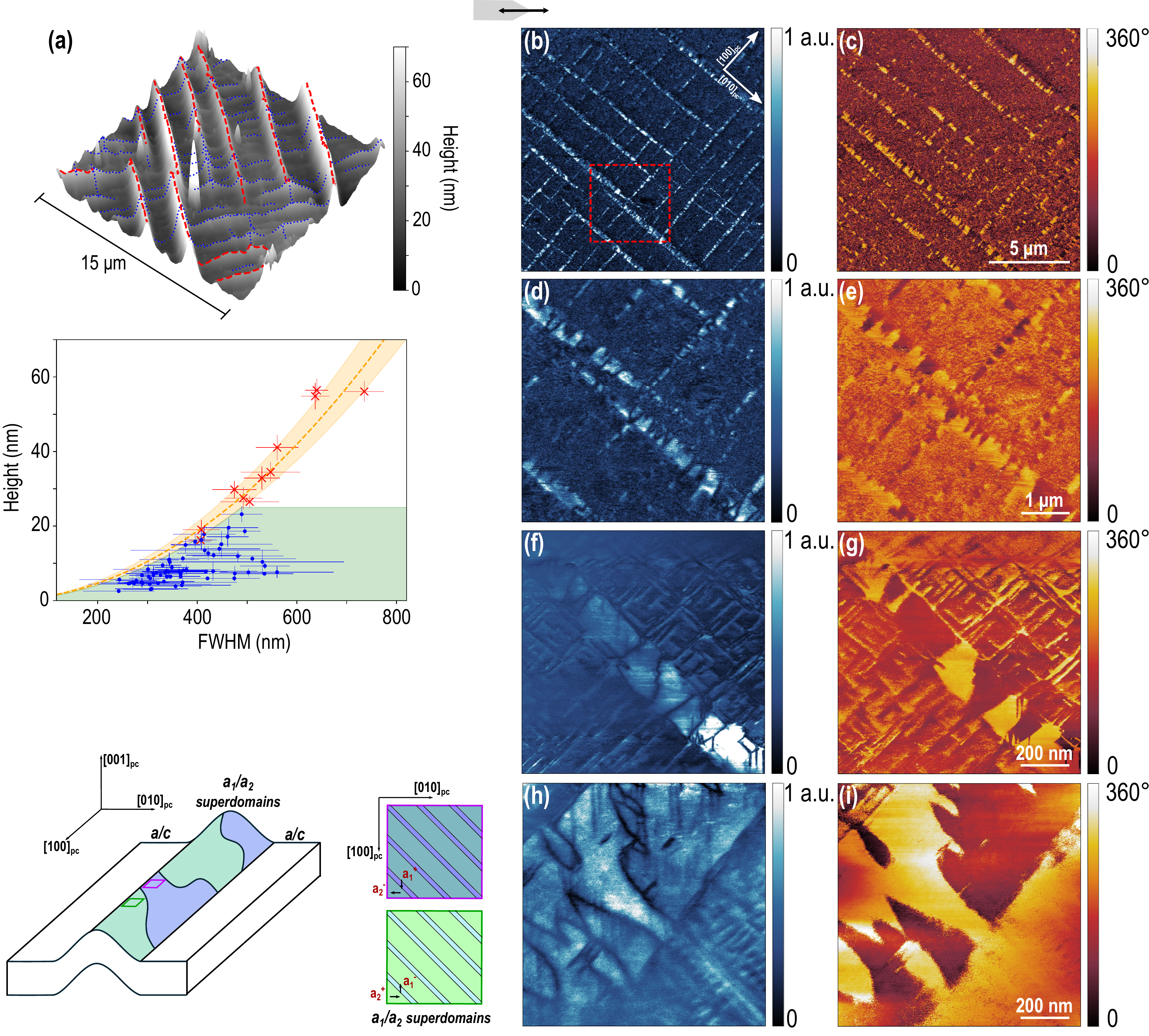}
  \caption{Classification of the ripples by height and FWHM, and lateral piezoresponse force microscopy analysis of the ripple pattern at different scales. (a) 3D AFM topography image of the ripples over a 15$\times$15 $\mu$m$^2$ area. Just below, the 2D scatter plot classifies the ripples based on their height and FWHM. (b, c) Corresponding PFM amplitude and phase signals of the area shown in (a). (d, e) PFM amplitude and phase signals of the region highlighted by the red dashed box in (b). (f, g) PFM amplitude and phase signals of a ``secondary'' ripple with height and FWHM of 7 nm and 300 nm, respectively. A purely in-plane $a$ domain configuration is observed at the ripple, while an $a/c$ domain structure is seen elsewhere. (h, i) PFM amplitude and phase signal of a ``primary'' ripple with height $\geq$ 100 nm and FWHM $\geq$ 1.5 $\mu$m, showing the presence of $a_1$ and $a_2$ domains organized in superdomains. The schematic on the left shows the possible in-plane domain configuration within the superdomains.}
\label{fig:PFM_wrinkles}
\end{figure}

Next, the ferroelectric properties were investigated at a smaller scale using higher-reso\-lu\-tion techniques. 
Piezoresponse force microscopy (PFM) was used to probe the local domain structure of both the ripples and the flat regions of the membrane. 
A preliminary PFM analysis was performed on the epitaxial heterostructure prior to lift-off and transfer, indicating an $a/c$ domain structure, as shown in SI Figure S6. 
PFM measurements of the membrane are reported in \textbf{Figure \ref{fig:PFM_wrinkles}}.
The topography image in Figure \ref{fig:PFM_wrinkles}a reveals two distinct types of ripples. The biggest ripples, designated as ``primary'' ripples, present large heights (up to 60 nm-high) that increase with the square of their full-width at half maximum (FWHM). These primary ripples show a characteristic radius of curvature of $\sim$ 1.5 $\mu$m at their crest. The ``secondary'' ripples, on the other hand, branch off from primary ripples or other secondary ripples and present both lower heights ($<$ 25 $\mu$m) and a larger FWHM distribution, which result in much larger radii of curvature (up to 8 $\mu$m).
The 2D scatter plot in Figure \ref{fig:PFM_wrinkles}a illustrates the two different ripple groups, classified according to their height and FWHM. Further details on the procedure used to determine the ripple parameters, as well as the computed values for a few selected examples, are provided in the SI section 2 and in SI Figure S7 and Figure S8.

Figures \ref{fig:PFM_wrinkles}b and \ref{fig:PFM_wrinkles}c display the lateral PFM amplitude and phase signals, respectively, obtained from the region shown in Figure \ref{fig:PFM_wrinkles}a. Figures \ref{fig:PFM_wrinkles}d and \ref{fig:PFM_wrinkles}e provide 5$\times$5 $\mu$m$^2$ lateral PFM amplitude and phase images acquired from the area marked by the red dashed box in Figure \ref{fig:PFM_wrinkles}b. Finally, Figures \ref{fig:PFM_wrinkles}f and \ref{fig:PFM_wrinkles}g show the amplitude and phase signals recorded on a ``secondary'' ripple, with a height and FWHM of approximately 7 nm and 300 nm, respectively. 

These measurements highlight the distinct contrast in domain configuration between the flat regions and the ripples. In the flat areas, the membrane retains an $a/c$ domain structure, consistent with the configuration observed prior to the lift-off and transfer process (see SI Figure S6 and Figure S9).
At the top of the ripples, the amplitude signals reveal the presence of two types of domain walls (appearing as minima, dark-colored lines): 180$^\circ$ domain walls of irregular shape and 90$^\circ$ domain walls oriented along the \{110\}$_{pc}$ directions. 
The 180$^\circ$ domain walls separate in-plane superdomains with opposite polarization directions, a conclusion supported by the phase signal, in which the yellow and red regions show a phase contrast of 180$^\circ$.
Meanwhile, the 90$^\circ$ domain walls, oriented along the \{110\}$_{pc}$ directions, correspond to the boundaries between the $a_1$ and $a_2$ domains, in which the polarization is rotated by 90$^\circ$. SI Figure S10 provides a schematic representation of the domain configuration at the ripple shown in Figures \ref{fig:PFM_wrinkles}f and \ref{fig:PFM_wrinkles}g, underlining the different types of domain walls.
To summarize, we observe the formation of superdomains at the ripples, composed of $a_1$ and $a_2$ domains.
In addition, at the edge of the ripple, stripe-like in-plane domains are observed in agreement with previous reports \cite{pesquera2024hierarchicaldomainstructuresbuckled}.

The presence of superdomains is further confirmed in Figure \ref{fig:PFM_wrinkles}h and Figure \ref{fig:PFM_wrinkles}i, which display the amplitude and phase signals of a larger ripple compared to the one of Figures \ref{fig:PFM_wrinkles}f and \ref{fig:PFM_wrinkles}g. These ripples that belong to the ``primary'' type, with heights reaching up to 100 nm and FWHM values extending to 1.5 $\mu$m, were predominantly observed at the edges of the membrane flakes. They were analyzed to gain a clearer understanding of the in-plane domain configuration within the ripples.
The superdomains, characterized by 180$^\circ$ domain walls and a corresponding 180$^\circ$ phase contrast, are clearly evident and their internal domain structure  can be resolved. Based on the pseudocubic axis orientation, we propose the following scenario to explain these experimental observations: within the superdomains, large $a_2$ domains (oriented along the [010]$_{pc}$ direction) coexist with narrower $a_1$ domains (oriented along the [100]$_{pc}$ direction), with their polarizations rotated by 90$^\circ$ relative to each other.
The schematic on the left of Figure \ref{fig:PFM_wrinkles}h illustrates the arrangement of $a_1$ and $a_2$ domains within the superdomains, with black arrows indicating their respective polarization directions. It is clear that the $a_1$ and $a_2$ domains are present in different proportions, leading to a net polarization predominantly aligned with the polarization direction of the $a_2$ domains. A more detailed schematic that illustrates the polarization direction within the superdomains is provided in SI Figure S11.
The corresponding topography images for the amplitude signals shown in Figures \ref{fig:PFM_wrinkles}b, d, f and g are provided in SI Figures S12a, S12b, S12c and S12d.

\begin{figure}[ht!]
\centering
  \includegraphics[width=\textwidth]{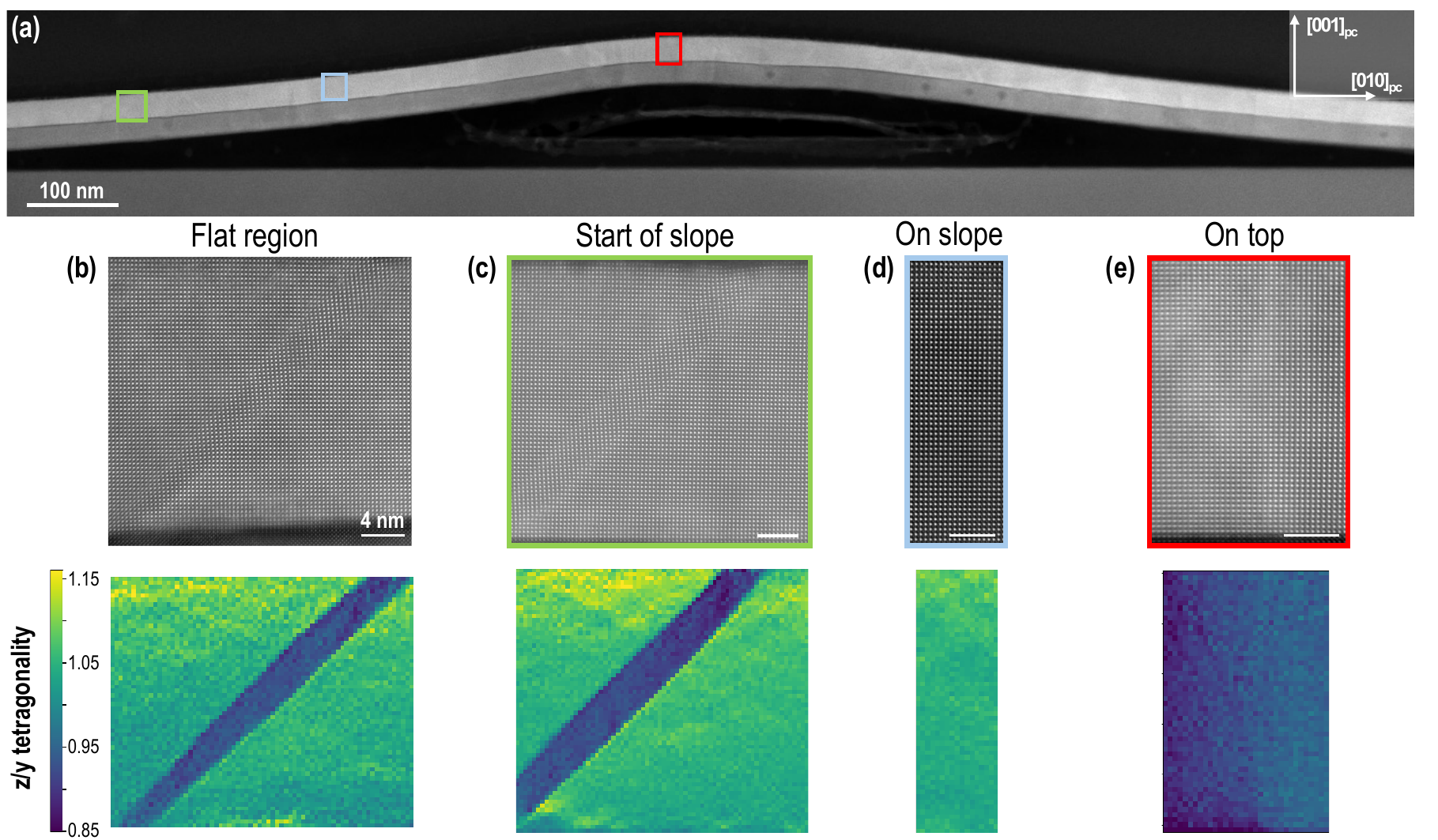}
  \caption{STEM cross-sectional study of a ``primary'' ripple (height$>$100 nm, FWHM$>$1$\mu$m). (a) STEM-HAADF image at low magnification showing the ripple. Image stacks were acquired and aligned in the highlighted regions, marked in different colors. (b) HAADF image and corresponding tetragonality analysis from a flat region far from the ripple, revealing an $a/c$ domain structure (for $c$ domain: $z/y>$1, for $a$ domain: $z/y\leq$1). (c) HAADF image and tetragonality analysis from a region highlighted in green, also showing an $a/c$ domain structure. (d) HAADF image and tetragonality analysis ($z/y>$1) from the slope of the ripple, indicating a pure $c$ domain structure (region highlighted in blue). (e) HAADF image and tetragonality analysis ($z/y<$1) from the summit of the ripple, revealing a pure $a$ domain structure (region highlighted in red).}
\label{fig:big_wrinkle_tem}
\end{figure}

To complete our findings regarding the domain configuration based on the top-view PFM measurements, and to gain deeper insights into the domain structure along the ripple profile, atomic resolution STEM-High Angle Annular Dark Field (HAADF) imaging was performed along a cut perpendicular to the ripple direction, as shown in \textbf{Figure \ref{fig:big_wrinkle_tem}a}. This provides a cross-sectional view of a very large ripple, which belongs to the ``primary'' type. Multiple atomic resolution image stacks were acquired from a flat region towards the top of the ripple. The domain structure was analyzed at different regions of interest, details of which are shown in Figures \ref{fig:big_wrinkle_tem}b, \ref{fig:big_wrinkle_tem}c, \ref{fig:big_wrinkle_tem}d, and \ref{fig:big_wrinkle_tem}e.
The STEM-HAADF images reveal that an $a/c$ domain configuration predominates in the flat regions, consistent with PFM observations, and extends to the initial part of the ripple slope (green-highlighted region in Figure \ref{fig:big_wrinkle_tem}a). A custom Python code based on Atomap \cite{nord_atomap_2017} was used to identify the atom positions in the STEM-HAADF images, enabling the calculation of the tetragonality ($z/y$, the ratio between the out-of-plane and in-plane lattice parameters).

Figures \ref{fig:big_wrinkle_tem}b and \ref{fig:big_wrinkle_tem}c confirm the presence of an $a/c$ domain configuration in both the flat region and the lower slope. In the tetragonality map, the $a$ domains appear as stripe-like blue regions where the $z/y$ ratio is less than 1, whereas $c$ domains correspond to green regions where the $z/y$ ratio is greater than 1. As the top of the ripple is approached, the domain structure transitions to a purely out-of-plane $c$ domain state (light blue-highlighted region in Figure \ref{fig:big_wrinkle_tem}a), as shown in Figure \ref{fig:big_wrinkle_tem}d, where all unit cells exhibit a $z/y$ ratio greater than 1.
At the summit of the ripple (red-highlighted region in Figure \ref{fig:big_wrinkle_tem}a), the polarization becomes purely in-plane, corresponding to an $a$ domain state. This is evident in Figure \ref{fig:big_wrinkle_tem}e, where the tetragonality analysis shows a $z/y$ ratio entirely below 1, consistent with the PFM results in Figure \ref{fig:PFM_wrinkles}. The appearance of regions where the polarization develops out-of-plane can be attributed to the negative curvature of these areas, which induces compressive strain, promoting an out-of-plane polarization. Conversely, at the top of the ripple, where the strain is tensile, the polarization aligns in-plane.

\begin{figure}[ht!]
\centering
  \includegraphics[width=\textwidth]{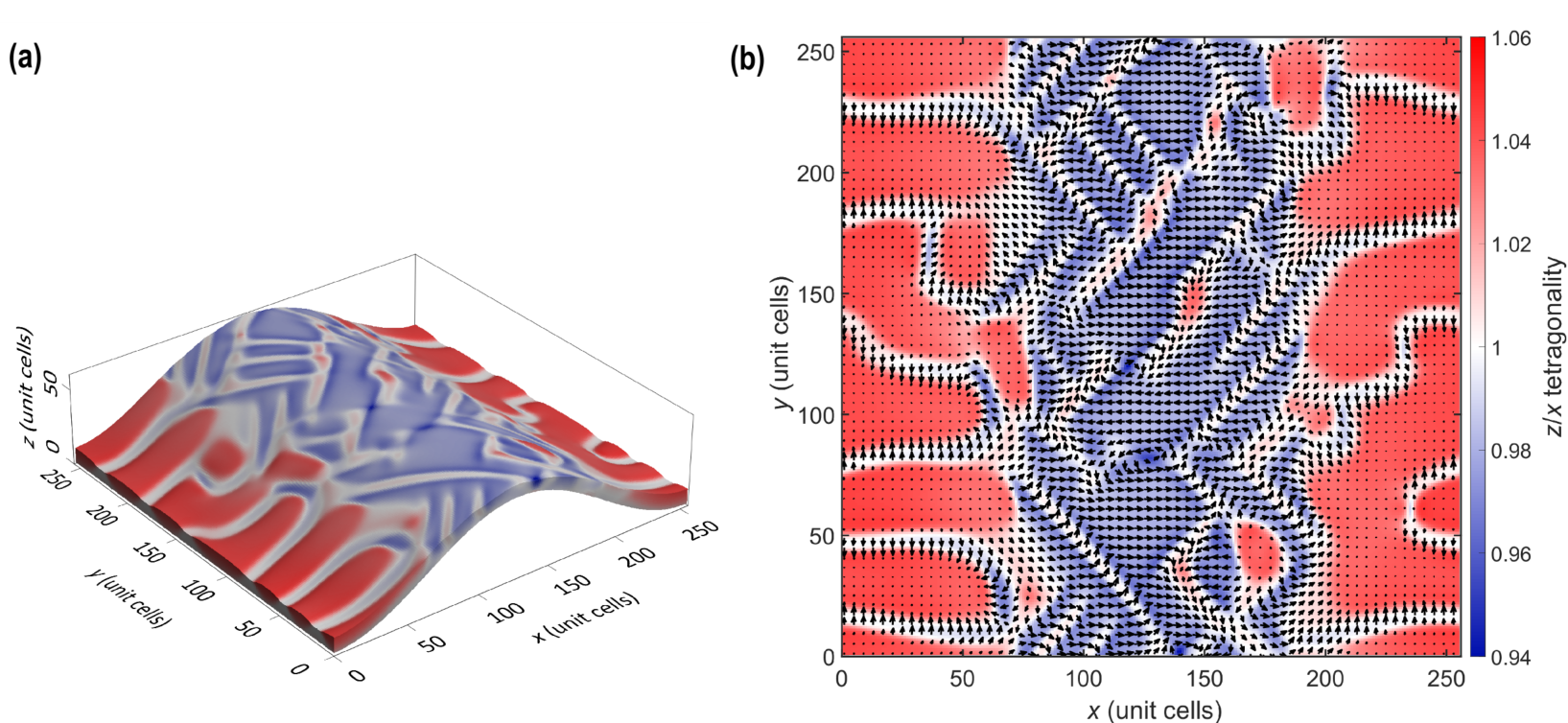}
  \caption{Domain structure naturally developed in a phase-field simulation of a thin film of PbTiO$_3$ subjected to an imposed curvature profile. (a) Overall view of the simulated system revealing both top and side views of the resulting domain structure as well as the imposed z-shift of the bottom part of the film (for clarity the shift is enhanced by a factor of 10). (b) Top view of the same domain structure. The color scale represents the strain ratio (1+$e_{zz}$)/(1+$e_{xx}$) labelled as $z/x$ tetragonality, the arrows in (b) are proportional to the in-plane polarization. The simulation box contains 256 $\times$ 256 $\times$ 8 individual points with a spacing of 0.4 nm.}
\label{fig:Phase_Field}
\end{figure}

To acquire a deeper understanding of the influence of the thin film curvature on its domain structure, phase-field simulations were performed using the Ginzburg-Landau model for ambient-temperature PbTiO$_3$, implemented in the FERRODO code \cite{Hlinka2006,Marton2006}, as recently applied in superlattice studies \cite{Zatterin2024,Hadjimichael2021}. Details of the parameters used for the phase-field simulations can be found in the Experimental section. 
The model incorporates a Landau potential expanded up to the 6$^{th}$-order terms in polarization \cite{Haun1987}, along with gradient, elastic, electrostrictive, and dipole-dipole long-range interactions, taking into account the background lattice permittivity and the screening of top and bottom surface charges.
Given the pseudocubic lattice parameters of SrRuO$_3$ and the observed residual $a$ domains in the flat region of the PbTiO$_3$/SrRuO$_3$ membrane, it is assumed that the macroscopic in-plane strain is nearly zero or slightly tensile relative to the cubic PbTiO$_3$ reference, and is set to 0.005 in both in-plane directions.
The simulation box with dimensions 256 $\times$ 256 $\times$ 8 represents a square-like section of the thin PbTiO$_3$ film with dimensions 2$L$ $\times$ 2$L$ $\times$ $d$, where $d$ is the thickness of the film. Periodic boundary conditions are applied in the $x$ and $y$ directions. 
A similar curvature geometry as documented in Figure \ref{fig:big_wrinkle_tem}, specifically a ripple extended along the substrate $y$//[010]$_{pc}$ axis, was introduced to the bottom surface as a $h \times \sin\left(\frac{\pi x}{2L}\right)^2$ profile. The height $h$ and the FWHM of the ripple profile $L$ were maintained fixed. The top surface was mechanically free.
\textbf{Figure \ref{fig:Phase_Field}} shows the relaxed domain structure obtained from the initial random white noise configuration after more than 200'000 iterations. The results show that the polarization is mainly out-of-plane on the side of the ripple, and predominantly in-plane at the top of the ripple. We notice that the sides of the ripple display an $a/c$ domain pattern with prevailing $c$ domains alternating with a small fraction of $a$ domains (about 10-20 \%, preferably a$_2$). In contrast, the top of the ripple displays superdomains formed by an $a_1/a_2$ domain pattern, with prevailing $a_1$ domain states, in good agreement with the SHG, PFM and STEM observations.

Since this phase-field code uses realistic gradient terms, the used approach is optimized for simulations with a spatial step of 0.4 nm, a value corresponding to the perovskite lattice parameter. Therefore, the real thickness of the simulated thin film shown in Figure \ref{fig:Phase_Field} is $d$ =3.2 nm, the height of the ripple is $h$=2.04 nm, and its FWHM is $L$ =51.2 nm. Nevertheless, the overall contribution of the gradient terms to the relaxed state energy is minor, so that it can be guessed that the main characteristics of the simulated domain structure do not depend significantly on the absolute length scale. To verify the conjecture that the domain pattern is primarily defined by the $L:d:h$ ratio, the domain pattern was rescaled from Figure \ref{fig:Phase_Field} by a factor of 10 by setting the spatial step to 4 nm. 
This corresponds to a 20.4 nm-high ripple with a FWHM of 512 nm, and a film thickness of 32 nm. More than 20'000 additional iterations of phase-field annealing with the same model were carried out, and, as anticipated, the domain pattern remained essentially the same indicating that the only relevant independent length scales in this problem are $L$, $d$, and $h$.

\begin{figure}[ht!]
\centering
  \includegraphics[scale=0.45]{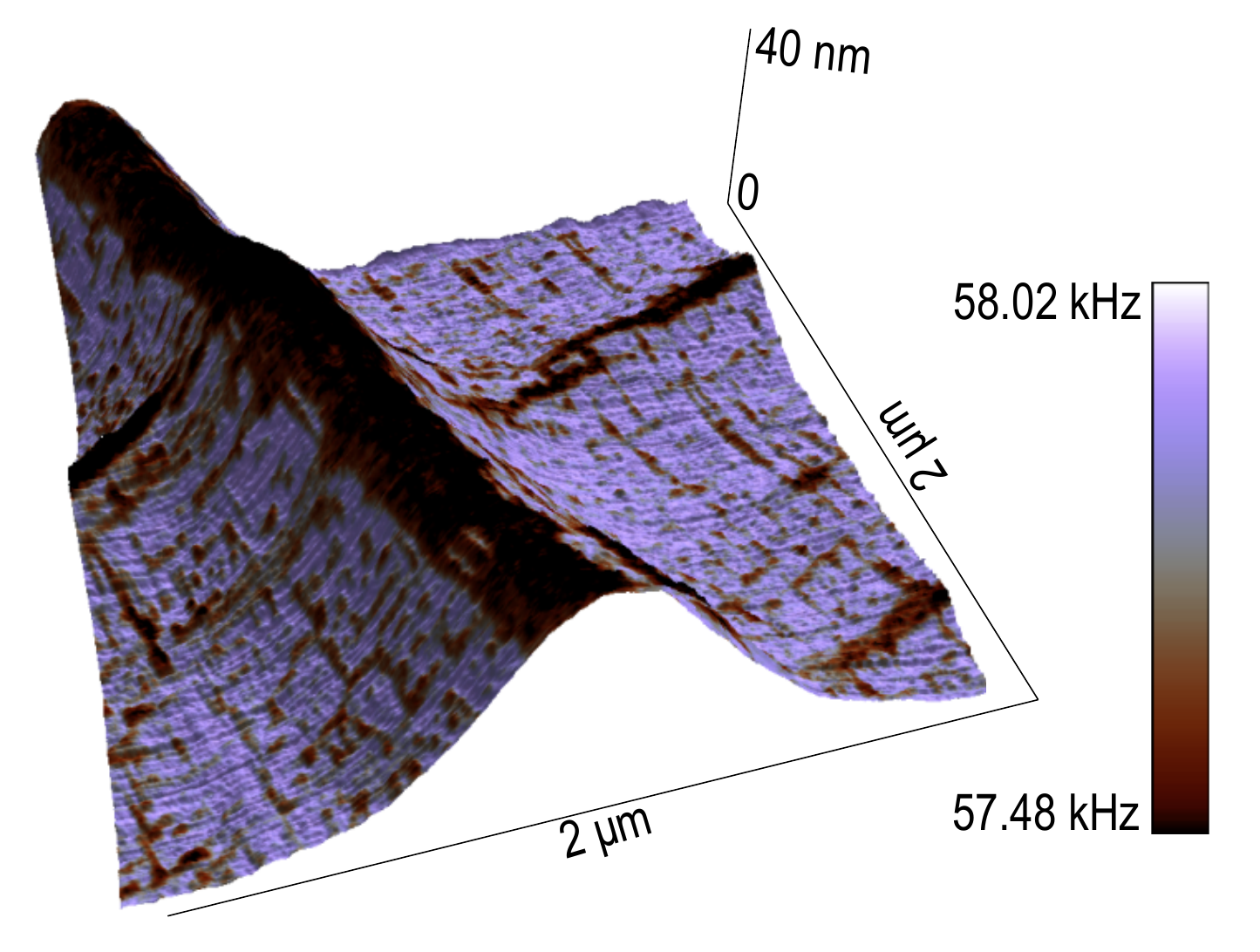}
  \caption{Mechanical properties of the PbTiO$_3$ ripple structures. Contact-resonance frequency image of a 40 nm-high ripple, revealing the mechanical stiffness distribution, with softer regions visible in both the ripple and flat areas.}
\label{fig:CR_AFM}
\end{figure}

After establishing the ferroelectric domain configuration of the ripples, the mechanical properties of the PbTiO$_3$/SrRuO$_3$ membrane were investigated using contact resonance force microscopy (CRFM), reported in \textbf{Figure \ref{fig:CR_AFM}}, which provides nanoscale insights into the material elasticity \cite{RABE2000430, 10.1063/5.0059930}. 
In this mode, the AFM tip maintains stable contact with the sample while the entire system undergoes mechanical excitation \cite{RABE2000430}. The tracked contact resonance frequency directly correlates with the Young modulus of the sample: lower frequencies indicate softer materials, while higher frequencies suggest stiffer ones. 
Figure \ref{fig:CR_AFM} shows the CRFM data overlaid on the 3D topography of a cross-ripple region. This region features a ``primary''-type ripple with a height of approximately 40 nm, and reveals the mechanical stiffness distribution across the membrane. Complementary vertical and lateral PFM signals, with 2D topography and CRFM images at the same location are shown in SI Figure S13.

Notably, the ripples exhibit distinct mechanical properties, with overall lower stiffness than the flat regions, and with in-ripple $a$ domains showing even softer characteristics compared to the surrounding areas. Regions of lower stiffness in the flat region, which align with the $a$ domains identified in the PFM measurements, were also detected. This observation suggests a correlation between polarization orientation and mechanical properties, a phenomenon consistent with previous studies on polarization-dependent mechanical and tribological asymmetry in ferroelectrics \cite{10.1063/5.0059930, Cho2024-uq}. 

\bigskip

To understand the mechanisms underlying the formation of the ripple pattern, it is important to clarify that the origin of these ripples is not attributed to the lift-off and transfer process or the properties of the target substrate onto which the membrane is transferred. Specifically, we find that the structural properties of the target substrate do not influence the formation or arrangement of the ripples. This is supported by the presence of a nanometer-scale gap between the membrane and the substrate, which prevents mechanical coupling and ensures that the structural properties of the membrane remain independent of the target substrate \cite{Segantini2024}.
The ripple formation is instead attributed to changes in electrostatic and mechanical boundary conditions following the release of the membrane from its growth substrate, as reported in previous studies \cite{Li2022}. 

We hypothesize that, since the screening effect provided by the SrRuO$_3$ electrode does not fully compensate for the bound polarization charges, after being released, the membrane that is now free from the substrate constraints reorganizes its domain configuration to minimize the electrostatic energy. This results in a partial in-plane re-orientation of the PbTiO$_3$ polarization. Additionally, the lattice mismatch between PbTiO$_3$ (tetragonal with $a=b$= 3.904 $\AA$, $c$=4.152 $\AA$) and SrRuO$_3$ (orthorhombic structure with $a_o$=5.57 $\AA$, $b_o$= 5.53 $\AA$, $c_o$= 7.85 $\AA$, corresponding to the pseudocubic unit cell parameters $a_{pc}=c_{pc}$= 3.924 $\AA$, $b_{pc}$=3.925 $\AA$) induces the bending of the membrane to further reduce elastic energy.

Furthermore, we propose that, although Figure \ref{fig:PFM_wrinkles}a suggests that large ripples (``primary'-type), which extend for tens of micrometers, preferentially develop along the [010]$_{pc}$ axis, the cubic symmetry of the SrTiO$_3$ substrate implies equivalent strain release along both the [100]$_{pc}$ and [010]$_{pc}$ directions. Consequently, large ripples should form along both crystallographic axes in equal proportion, as these directions offer the highest strain-relief potential. Smaller ripples, on the other hand, develop perpendicular to the larger ones and are confined by them to accommodate the residual strain.

The geometry of the ripples introduces distinct strain conditions compared to the flat regions of the membrane, leading to differences in domain configurations. The SHG and PFM analyses reveal that the polarization at the top of the ripples is purely in-plane, with in-plane domains organizing into superdomains composed of varying proportions of $a_1$ and $a_2$ domains, consistent with the phase-field model. Notably, the unequal volume fraction of $a_1$ and $a_2$ domains within the $a_1/a_2$ superdomains can increase the strain, reaching a limit similar to that of a pure $a$ domain with polarization aligned along the direction tangential to the ripple curvature. This natural limit corresponds to the observed upper bound of curvature ( $\sim$ 1.55 $\mu$m), as documented in SI Figure S7b. 
Using the classical Timoshenko bi-metal curvature formula \cite{timoshenko1925analysis} for the PbTiO$_3$/SrRuO$_3$ bilayer, with a thickness ratio of approximately 1.5 and a Young modulus ratio between 1 and 6, the 1.55 $\mu$m curvature implies a misfit strain of only about 2.5-3.5 \%. This is roughly half the nominal misfit between the $c$ lattice parameter of bulk stress-free PbTiO$_3$ and the lattice parameter of bulk SrRuO$_3$, suggesting that in addition to the curvature, the bilayer is uniformly stretched. Consequently, the polarization and tetragonality of PbTiO$_3$ are likely smaller than in the bulk, probably due to the incomplete screening of the bound charges. Surface tension effects could also play a role in reducing the observed curvature.
The varying proportions of $a_1$ and $a_2$ domains likely account for the needle-like shape of the ripples, which do not extend throughout the entire membrane flake, as shown in SI Figures S1b and S12a.
Moreover, the PFM analysis presented in Figures \ref{fig:PFM_wrinkles}h and \ref{fig:PFM_wrinkles}i shows that, when the ripple forms along the [100]$_{pc}$ axis, the superdomains are predominantly composed of $a_2$ domains. In contrast, the phase-field simulations (Figure \ref{fig:Phase_Field}) suggest that when the ripple develops along the [010]$_{pc}$ direction, $a_1$ domains are more abundant. This indicates that the relative proportions of $a_1$ and $a_2$ domains within the ripples are orientation-dependent.

Finally, the ripple pattern that forms upon the membrane detachment is highly dependent on the thickness of both the SrRuO$_3$ bottom electrode and the PbTiO$_3$ layer, as it was reported in a previous work \cite{Huang2024}. Different thickness configurations are expected to lead to variations in the electrostatic and mechanical interactions, thereby affecting the period, height, and FWHM of the ripples.


In summary, in this work we have demonstrated that releasing a ferroelectric material from its growth substrate leads to the formation of a self-organized, adaptive ripple pattern. This pattern arises from the interplay between mechanical constraints and electrostatic effects associated with detachment and the presence of a back electrode. The observed ripples develop along the pseudocubic axes of PbTiO$_3$ and can be classified based on their height and FWHM.
The nonlinear optical measurements revealed that the structural inversion symmetry breaking is systematically controlled by the ripple geometry. At the crests of the ripples, the polarization is fully in-plane, in contrast to the domain configuration in the flat regions, and ripples aligned along specific pseudocubic directions induce a rotation of the polarization.
Additionally, ripples exhibit distinct mechanical properties compared to the flat areas. Notably, the presence of softer $a$ domains in the highly strained regions of the ripple structure highlight the bidirectional coupling between structural deformation and polarization in ferroelectric materials. This interplay suggests that strain engineering not only can be used to modulate the polarization, but the polarization itself can influence the mechanical properties. Such tunability opens new avenues for applications where a controlled ripple formation enables polarization engineering, and conversely, polarization control can be leveraged to tailor mechanical responses, broadening the potential for novel device functionalities.
This study indicates that modifying the mechanical boundary conditions to which the membrane is subject --- by adjusting the electrode thickness, varying the ferroelectric material thickness, or removing the electrode altogether --- it is possible to tune strain and control the domain configuration. Consequently, this could enable control over the ripple pattern itself, including the density and dimensions of the ripples. Moreover, by tuning the domain configuration, the mechanical properties of the ripples can be adjusted, offering precise control over both structural and functional properties.
Finally, these findings could pave the way for novel device concepts that exploit rippled ferroelectrics as platforms for acoustic or optical wave confinement. Furthermore, when combined with the inherent flexibility of the membrane system, these structures could serve as building blocks for next-generation flexible electronic and optoelectronic devices.

\section{Experimental Section}
\subsection{Sample Preparation}
The 15 nm-thick Sr$_3$Al$_2$O$_6$ sacrificial layer was epitaxially grown on a SrTiO$_3$(001)-oriented, non-terminated substrate using pulsed laser deposition (PLD) equipped with a KrF excimer laser ($\lambda = 248$ nm). The Sr$_3$Al$_2$O$_6$ layer was deposited at a temperature of 700 $^\circ$C in an oxygen pressure of $10^{-5}$ mbar, using a laser fluency of 2.2 J·cm$^{-2}$ with a repetition rate of 1 Hz. The 4-unit-cell-thick SrTiO$_3$ capping layer was subsequently deposited by PLD under the same temperature and oxygen pressure conditions, using a laser fluency of 1.6 J·cm$^{-2}$. After deposition, the sample was cooled down in the same oxygen pressure for 180 minutes.
The capping layer was used to prevent the degradation of Sr$_3$Al$_2$O$_6$ upon exposure to air.
The 22 nm-thick SrRuO$_3$ layer and the 30 nm-thick PbTiO$_3$ layer were grown using in-house constructed off-axis radio-frequency magnetron sputtering system. The SrRuO$_3$ layer was deposited from a stoichiometric target at 660$^\circ$C in 100 mTorr of O$_2$/Ar mixture of ratio 4:80, at a power of 80 W. The PbTiO$_3$ layer was deposited at 560$^\circ$C, in 180 mTorr of a 20:29 O$_2$/Ar mixture, at a power of 60 W, and using a Pb$_{1.1}$TiO$_3$ target with 10\% excess of Pb to compensate for its volatility.
The lift-off procedure was performed by applying a polydimethylsiloxane (PDMS) strip to the PbTiO$_3$ surface, followed by transferring the structure to de-ionized water for approximately 1 hour to dissolve the Sr$_{3}$Al$_{2}$O$_{6}$ sacrificial layer. Subsequently, the PDMS with the resulting PbTiO$_3$/SrRuO$_3$/SrTiO$_3$(capping layer) membrane was transferred onto the Nb:SrTiO${_3}$(001) (0.5 wt\%) non-terminated substrate. The transfer process consisted in placing the Nb:SrTiO$_3$(001) substrate on a hot plate maintained at 80 $^\circ$C. Once the PDMS/membrane system adhered to the substrate, it was pressed down for 90 seconds.
\subsection{Optical Second Harmonic Generation (SHG)}
In the SHG experiment, a femtosecond light pulse with a central wavelength of 800 nm and a repetition rate of 1 kHz was irradiated, and the SHG wave with a central wavelength of 400 nm was monitored using a photomultiplier tube (PMT). To achieve micrometer-sized spatial resolution, the fundamental wave was focused to the diffraction limit on the sample surface using a $\times$100 objective lens with a Numerical Aperture (N.A.) of 0.8. The fundamental and SHG waves were spectrally filtered using band-pass filters (BPF) and short-pass filters (SPF), respectively. The polarization states of the fundamental and SHG waves were controlled using a half-wave plate (HWP) and a Glan-Talyer polarizer, respectively.
\subsection{Piezoresponse Force Microscopy (PFM)}
Piezoresponse force microscopy measurements were performed under ambient conditions using a commercial Asylum Research MFP-3D atomic force microscope in Dual AC Resonance Tracking (DART) mode, by applying a driving amplitude of 500 mV at high-frequency, approximately 60 kHz for vertical PFM and 160 kHz for lateral PFM. A Bruker CONTV-PT cantilever with a nominal spring constant of 0.2 N/m and resonance frequency of 13 kHz was employed. Silver paste was applied to a corner of the sample in contact with the bottom SrRuO$_3$ electrode close to the region being scanned in order to establish a well-defined reference for the applied bias.
\subsection{Scanning Transmission Electron Microscopy (STEM)}
The TEM lamella in this study were prepared by focused ion beam milling using a Zeiss CrossBeam 540. Ripples of different sizes can be easily identified before preparation. A large ripple, with height $\geq$ 100 nm, was targeted for this study. The lamella preparation follows a typical preparation procedure with milling beam energy of 30 keV and currents varying from 1.5 nA down to 80 pA. At the end of milling, each lamella was cleaned using a 5 keV beam with a current of 30 pA. Afterwards, the lamella were further cleaned using the Gatan precision ion polishing system 2 at 0.5 kV for about 20 minutes, which removes the artefacts induced during Ga milling and further thins down the lamella. 
The STEM analyses were carried out on a double aberration-corrected FEI Titan Themis 80-300, using a 300 keV beam energy, 20 mrad probe convergence semi-angle, and with beam currents down to $\sim$40 pA. A Fischione HAADF detector was used, with camera length adjusted to give a nominal collection semi-angle of 50.5–200 mrad. STEM-HAADF images are recorded as 90$^{\circ}$ rotation series, which are then processed using the Smart Align plug-in in DigitalMicrograph3 to perform rigid and non-rigid alignment to maximize measurement precision \cite{jones_smart_2015}. 
The magnification of each image stack was carefully selected to ensure that most of the PbTiO$_3$/SrRuO$_3$ membrane was captured while maintaining sufficient pixel density per atom for atom position identification. Recording the series with up to 4k pixel frame sizes allowed for high precision across films tens of nanometers thick.
A home-made python code based on Atomap \cite{nord_atomap_2017} was used to analyze the tetragonality in the PbTiO$_3$ within the PbTiO$_3$/SrRuO$_3$ membrane. The code is available on github (https://github.com/PauTorru/ucsplit). First, the entire image was sectioned into unit cells, afterwards, the four A cation positions could be identified in each unit cell. Once the atom positions were refined, the tetragonality ($z/y$ ratio) could be easily calculated for each unit cell in a given HAADF image. The $z/y$ ratio greater than 1 represents the $c$ domain state, while $z/y$ ratio lower than 1 represents the $a$ domain state.
\subsection{Phase-field Simulations}
Phase-field simulations of PbTiO$_3$ films were carried out using the program FERRODO \cite{Hlinka2006,Marton2006}, which allows to find stationary domain structure configurations in perovskite ferroelectrics defined by the generalized Ginzburg–Landau–Devonshire model \cite{Domain_walls_from_fund}. Ferroelectric PbTiO$_3$ was described by the same model as in References \cite{Hadjimichael2021,Zatterin2024}.
The Landau parameters at 298 K are the following: $\alpha_1=-1.709 \times 10^8 \ JmC^{-2}$, $\alpha_{11}=-7.25 \times 10^7 \ Jm^5C^{-4}$, $\alpha_{12}= 7.5 \times 10^8 \ Jm^5C^{-4}$,  $\alpha_{111}=2.61 \times 10^8 \ Jm^9C^{-6}$, $\alpha_{112}=6.1 \times 10^8 \ Jm^9C^{-6}$, $\alpha_{123}=-3.66 \times 10^9 \ Jm^9C^{-6}$, with gradient parameters, $G_{11}=1\times 10^{-10} Jm^3C^{-2}$, $G_{12}=-1\times 10^{-10} Jm^3C^{-2}$, $G_{44}=1\times 10^{-10} Jm^3C^{-2}$, elastic components $C_{11}=1.746\times 10^{11} Jm^{-3}$, $C_{12}=7.94\times 10^{01} Jm^{-3}$, $C_{44}=1.111\times 10^{11} Jm^{-3}$, and electrostriction parameters $q_{11}=1.1412 \times 10^{10} JmC^{-2}$, $q_{12}=4.6 \times 10^{8} JmC^{-2}$,  $q_{44}=7.5 \times 10^{9} JmC^{-2}$, $Q_{11}=0.089 \ m^{4}C^{-2}$, $Q_{12}=-0.026 \ m^{4}C^{-2}$, $Q_{44}=0.0675 \ m^{4}C^{-2}$. The background permittivity $\epsilon_B$, as defined in References \cite{Hlinka2006,Marton2006}, was set to 1. The dynamics of the system is considered purely dissipative, described by the Landau-Khalatnikov equation. Simulations are conducted in real space, which allows us to prescribe the local shape of the sample, as well as treat the bottom and top surfaces of the film independently. The electrostatics and the dipole-dipole interaction is taken into account by explicitly solving the Poisson's equation for electric potential, considering the local polarization bound-charge density under proper boundary conditions (in this case shorted) on both the top and bottom surfaces. The local electric field was then evaluated from the electrostatic potential, and the interaction between the local ferroelectric polarizations and this local electric field was incorporated.
\subsection{Contact-resonance Force Microscopy (CRFM)}
Contact resonance force microscopy (CRFM) measurements were conducted under ambient conditions using a Cypher-ES commercial atomic force microscope (Asylum Research). A CONTV-PT probe (Bruker) with a nominal spring constant of 0.2 N/m and resonance frequency of 13 kHz was employed. The selection of a soft cantilever was crucial to minimize potential deformation of the membrane and ripple structures during imaging. Despite the low spring constant, a stable contact between the AFM tip and the sample was maintained throughout the measurements. In CRFM mode, the entire system undergoes mechanical excitation while the AFM tip remains in stable contact with the sample surface. The applied force is regulated to remain constant during the measurement process. This approach allows for high-resolution mapping of mechanical properties while minimizing sample damage, making it particularly suitable for investigating delicate structures such as ripples within the membrane.

\medskip
\textbf{Acknowledgements} \par 
This work was supported by the Swiss State Secretariat for Education, Research and Innovation (SERI) under contract no. MB22.00071, the Gordon and Betty Moore Foundation (grant no. 332 GBMF10451 to A.D.C.), the European Research Council (ERC), by the Dutch Research Council (NWO) as part of the VIDI (project 016.Vidi.189.061 to A.D.C.), the ENW-GROOT (project TOPCORE) programmes and by the Swiss National Science Foundation – division II (200020 207338). We acknowledge the Interdisciplinary Centre for Electron Microscopy (CIME) at EPFL for providing access to their electron microscopy facilities.
L.T. and C.L. acknowledge support by Division II of the Swiss National Science Foundation under project 200021\_200636. 
S.C. acknowledges the support from the Basic Science Research Program through the National Research Foundation of Korea (NRF), funded by the Ministry of Education
(RS-2024-00413670).
J. H.,  P. M. and P. O. acknowledge the assistance provided by the Operational Programme Johannes Amos Comenius of the Ministry of Education, Youth and Sport of the Czech Republic, within the frame of project Ferroic Multifunctionalities (FerrMion) [Project No. CZ.02.01.01/00/22\_008/0004591], co-funded by the European Union. The authors thank M. Matthiesen his support.

\medskip

%
\bibliographystyle{MSP}
\bibliography{bibliography}



\end{document}